\RequirePackage{fix-cm}
\documentclass[smallextended]{svjour3}       
\smartqed  
\usepackage{graphicx}
\usepackage{amsmath}
\newcommand{\be}{\begin{equation}}
\newcommand{\ee}{\end{equation}}
\newcommand{\eqnp}[1]{Eq.~(\ref{#1})}

\begin{document}

\title{Turbulence as a problem in non-equilibrium statistical mechanics 
}
\subtitle{For special issue of J. Stat. Phys. in memory of Leo P. Kadanoff}

\titlerunning{Nonequilibrium statistical mechanics of turbulence}        

\author{Nigel Goldenfeld         \and Hong-Yan Shih
}


\institute{Hong-Yan Shih and Nigel Goldenfeld \at
              Loomis Laboratory of Physics \\
              University of Illinois at Urbana-Champaign,\\
              1110 West Green Street,\\
              Urbana,\\
              IL 61801,\\
              USA\\
              \email{nigel@uiuc.edu}       \\
              \email{hshih7@illinois.edu}    
}

\date{Received: date / Accepted: date}

\maketitle

\begin{abstract}

The transitional and well-developed regimes of turbulent shear flows
exhibit a variety of remarkable scaling laws that are only now
beginning to be systematically studied and understood.  In the first
part of this article, we summarize recent progress in understanding the
friction factor of turbulent flows in rough pipes and
quasi-two-dimensional soap films, showing how the data obey a
two-parameter scaling law known as roughness-induced criticality, and
exhibit power-law scaling of friction factor with Reynolds number that
depends on the precise form of the nature of the turbulent cascade.
These results hint at a non-equilibrium fluctuation-dissipation
relation that applies to turbulent flows.  The second part of this
article concerns the lifetime statistics in smooth pipes around the
transition, showing how the remarkable super-exponential scaling with
Reynolds number reflects deep connections between large deviation
theory, extreme value statistics, directed percolation and the onset of
coexistence in predator-prey ecosystems. Both these phenomena reflect
the way in which turbulence can be fruitfully approached as a problem
in non-equilibrium statistical mechanics.

\keywords{Turbulence \and Phase Transitions \and Directed Percolation
\and Extreme Value Statistics \and Non-equilibrium statistical
mechanics \and Fluctuation-dissipation theorem \and Predator-prey ecosystems}

\PACS{64.60.ah \and 47.27.nf \and 02.50.-r \and 05.70.Ln}
\end{abstract}


\def\Re{\textrm{Re}}
\def\vK{von K\'arm\'an}
\def\vtr{\overline{u}_\theta(r)}

\section{Introduction}
\label{intro}

Fluid turbulence exhibits two regimes where universal scaling behavior
can be found.  The most studied of these is fully-developed turbulence,
which arises at asymptotically large Reynolds numbers, where there are
a host of scaling laws in a wide variety of different flows.  The
general idea is that these scaling laws are manifestations of some type
of critical point at infinite Reynolds number that controls scaling for
large and finite Reynolds numbers through what are essentially
crossover effects.  This perspective seems rather simple, but it
permits us to understand experimental data on turbulent pipe flows that
date back to 1933.

The other regime is the laminar-turbulence transition, which was first
studied scientifically by Reynolds in 1883 \cite{reynolds}.  Not until
the early 21st century were detailed and sufficiently systematic
measurements available to challenge and drive theoretical development.
The point here is that this transition is not to be regarded as an
outcome of low dimensional dynamical systems theory, but is in fact a
genuine non-equilibrium phase transition, exhibiting its own critical
point scaling laws that can be measured in experiment and calculated in
theory.  We will see in fact that this transition is most likely to be
in the universality class of directed percolation.

This article, in memory of Leo P. Kadanoff, describes selected
recent developments in these areas, from the unifying perspective that
turbulence should be approached as a problem in non-equilibrium
statistical mechanics. These examples demonstrate the utility of the
conceptual framework of non-equilibrium statistical mechanics applied
to turbulence, and suggests that there may be other fruitful extensions
to explore.  We are neither the first nor the only authors to have this
perspective; for example, see the book \cite{cardy2008non} or the work
of Ruelle, who applies this perspective to the problem of multi-fractal
scaling in turbulence \cite{ruelle2014non}.  However, the
examples presented here are centered around readily observable
phenomena that have not been previously considered in the framework of
non-equilibrium statistical mechanics, and make new predictions that
have been tested experimentally.  Leo Kadanoff himself was especially
interested in both phase transitions and turbulence, and some of his
most enduring contributions were in these areas.  The last detailed
conversation the authors held with Leo revolved around these topics,
and turned out to be influential in our subsequent work on these
topics.  Thus we are honored to have this opportunity to pay tribute to
his memory with this contribution.

\section{Friction factor of turbulent flow in rough pipes}
\label{sec:ff}

Fully-developed turbulence shares many features in common with critical
phenomena.  They are both characterized by strong fluctuations and
power-law scaling \cite{EYIN94}, and naively do not seem to possess a
small parameter that can be used to obtain perturbative results for the
difference between the mean field scaling exponents and those found in
experiment or numerical calculation.  Such things have been known in
the framework of turbulence since the time of Kraichnan, Edwards and
others
\cite{wyld1961formulation,kraichnan1959structure,edwards1964statistical,mccomb1995theory,yakhot1986renormalization,SREE05}.

It is therefore natural to ask why it is that the critical phenomenon
problem has been solved, but not the turbulence one.  To answer this,
we should recall how it was that critical phenomena came to be
understood, and what were the crucial steps.  The complex history of
this problem has been extensively reviewed by the active participants
\cite{kadanoff2009more,wilson1983renormalization,fisher1998renormalization,kadanoff2013relating,widom2011laboring,polyakov2015kenneth}
(see also \cite{cao1993conceptual} for historical context and the
relationship to renormalization in field theory), but the key steps can
be seen by going backwards in time.  The breakthrough in the problem is
the 1971 article by Wilson \cite{wilson1971renormalization}, whose very
title (\lq\lq Renormalization Group and the Kadanoff Scaling Picture"
--- a rare instance of a Physical Review editor allowing a title to
refer to an individual) indicates that the renormalization group
emerged from the key insights of Kadanoff's 1966 paper on the Ising
model \cite{kadanoff1966}.  Kadanoff's paper, in turn, opens by
summarizing Widom's discovery that the free energy of a system near its
critical point is a homogeneous function of the relevant coupling
constants \cite{WIDO65}.  Kadanoff went on to show how Widom's scaling
could arise by constructing the effective Hamiltonian at different
scales, and making certain technical assumptions. Although incapable of
computing critical exponents, the Kadanoff block spin picture, as it
came to be known, truly laid the basis for the complete renormalization
group solution to the phase transition problem.

The moral of this story for turbulence is that if we are to look for a
renormalization group style framework in which to understand
turbulence, the starting question should be: what is the analogue of
Widom's scaling law in turbulence?

\subsection{Widom scaling}
\label{sub:widom}
In the language of magnetic systems, Widom's scaling law is the
statement that the magnetization $M$ as a function of external field
$H$ and temperature $T$ is in fact a function of a single variable:
\begin{equation}
M(H,T) = |t|^{\beta} F(H/|t|^{\beta \delta})
\label{eqn:Widom}
\end{equation}
where $\beta$ and $\delta$ are critical exponents for the order
parameter and breakdown of linear response theory at the critical
isotherm respectively, reduced temperature $t\equiv (T-T_c)/T_c$ and
$F$ is a universal scaling function.  This data collapse formula is
equivalent to two asymptotic scaling laws near the critical point.  The
first is the order parameter scaling law
\begin{equation}
M \sim |t|^\beta
\label{eqn:op-scaling}
\end{equation}
for $H=0$ as $T\rightarrow T_c$.  The second is the breakdown of linear
response theory at the critical point.  Normally the induced
magnetization is proportional to the externally applied field for
sufficiently small $H$.  However at the critical point, this
relationship becomes a power-law with
\begin{equation}
M\sim H^{1/\delta}
\label{eqn:critical-isotherm}
\end{equation}
for $T=T_c$.  The data collapse formula Eq.~(\ref{eqn:Widom}) in effect
connects the macroscopic thermodynamics of the critical point with the
spatial correlations at small scales, as can be seen from using the
other scaling laws and the static susceptibility sum rule
\cite{goldenfeld1992lectures}. In order for Eq.~(\ref{eqn:Widom}) to be
equivalent to the two asymptotic scaling laws Eqs.
(\ref{eqn:op-scaling}) and (\ref{eqn:critical-isotherm}), the scaling
function $F(z)$ must be a particular power-law function of its argument
$z$ for large $z$, so that for $H\neq 0$ and $t\rightarrow 0$, the
vanishing $t$-dependent prefactor and the diverging $t$-dependent
argument of $F$ \lq\lq cancel out", leaving simply the power-law
function of $H$ that applies on the critical isotherm.

\subsection{Roughness-induced criticality: Widom scaling for
wall-bounded turbulence}
\label{sub:roughness}
To find an analogue for turbulent fluid flow in a pipe, we need to
first ask what is special about $T$ and $H$.  The reduced temperature
controls the distance to the critical point, and $H$ can be thought of
as a variable which couples to the degrees of freedom to bias them to
be ordered.  The turbulent analogue of $t$ could be taken to be the
inverse of the Reynolds number, Re.  The turbulent analogue of $H$
could be wall-roughness; the logic is that in a smooth pipe, the
laminar flow is linearly stable to all Re, but wall-roughness on a
scale $r$ can create disturbances that grow downstream and eventually
fill a pipe with turbulence.  If we are going to construct an analogue
of the critical point, we will need experimental data that
systematically cover as many decades of the control parameters
$\Re^{-1}$ and $r/D$ as possible, where the pipe diameter $D$ has been
introduced to non-dimensionalize the wall-roughness.

There is one experiment in the whole history of turbulence which
contains enough data to work with, and that is due to Nikuradze, who
was associated with Prandtl's laboratory during the 1930's
\cite{NIKU33}. Nikuradze's experiment, never repeated, extended to
$\Re\sim 10^6$, encompassing the crossover between laminar and
turbulent flows around $\Re\sim 2000$, where the change from Stokes
drag is marked by a sudden increase in drag over a small range of Re
around 1000-2000 (sometimes known as the \lq\lq drag catastrophe").
The experiment also covered one and a half decades in wall-roughness,
using the same pipe-flow geometry, and measured the normalized pressure
drop $\Delta P$ along the pipe as a function of Re and $r/D$.  The
pressure drop was normalized to yield the so-called friction factor
\begin{equation}
f\equiv \frac{\Delta P/L}{\rho U^2}
\label{eqn:friction-factor-def}
\end{equation}
where $L$ is the pipe length, $\rho$ is the fluid density and $U$ is
the mean flow speed.

Next we should ask about the analogue for the two asymptotic scaling
laws near the critical point.  The limit $T\rightarrow T_c$ is, from
our assumptions, equivalent to $1/\Re \rightarrow 0$, whereas the
$H\rightarrow 0$ limit is simply equivalent to $r/D\rightarrow 0$ where
$D$ is the diameter of the pipe.  The existence of this critical point
is sometimes known as \lq\lq roughness-induced criticality"
\cite{GOLD06}. The order parameter scaling law applies for $H=0$ and
thus corresponds to the behavior of the turbulent fluid as $r/D
\rightarrow 0$.  Nikuradze's experiments show that in the turbulent
regime, as the wall-roughness diminishes, the friction factor follows
further and further along the asymptote
\begin{equation}
f\sim \Re^{-1/4}.
\label{eqn:Blasius}
\end{equation}
This scaling was first observed by Blasius
\cite{blasius1913ahnlichkeitsgesetz}, and is the analogue of the order
parameter scaling law, Eq. (\ref{eqn:op-scaling}).  The critical
isotherm scaling representing the breakdown of linear response theory
corresponds to the behavior when $\Re\rightarrow\infty$.  In this
limit, the friction factor becomes independent of Re, and follows the
so-called Strickler scaling law \cite{STRI23}
\begin{equation}
f\sim (r/D)^{1/3}.
\label{eqn:Strickler}
\end{equation}

These stylized facts can be combined into a single scaling law,
following the same scaling calculation described above for magnets.
The result is that
\begin{equation}
f \left(\frac{r}{D}, \Re \right)=\Re^{-1/4} F \left(\frac{r}{D} \Re^{3/4} \right)
\label{eqn:data-collapse}
\end{equation}
where $F(z)$ is a universal scaling function whose asymptotic behavior
at small and large values of its argument are determined by the scaling
calculation.  This scaling law can be readily tested by replotting Nikuradze's data
in the form of $f \Re^{1/4}$ vs. $\Re^{3/4} \times r(/D)$, and a very
encouraging data collapse is found \cite{GOLD06}.  However, the
collapse is not perfect, and it is important to understand why.

\subsection{Anomalous dimensions in turbulence}
\label{sub:anom}

Turbulence, just as with critical phenomena, is characterized by
incomplete similarity \cite{barenblatt1972self,barenblatt1996scaling}.
This term, originally used in the context of similarity solutions to
deterministic partial differential equations, means that
self-similarity is weakly broken by a variable whose small value with
respect to the characteristic scale of the solution is nevertheless not
negligible.

In critical phenomena, for example, the correlation length $\xi (T)$
diverges near the critical temperature $T_c$, so that it becomes much
larger than the ultra-violet cut-off $\ell$, such as the lattice
spacing in solid state physics.  Even though $\ell/\xi(T) \rightarrow
0$ as $T\rightarrow T_c$, $\ell$ itself is not negligible, and in fact
it is this fact which is responsible for the existence of anomalous
critical exponents whose value differs from that expected on the basis
of mean field theory.  A clear example is the scaling of the two-point
correlation function $G(x-x', T_c)$ at the critical point: on
dimensional grounds, its Fourier transform $\hat G(k,T_c)$  has to have
dimensions of (length)$^2$ so that one would expect that $\hat G(k,
T_c) \sim k^{-2}$.  In fact, the scaling obeys $\hat G(k, T_c) \sim
k^{-2 + \eta}$, where $\eta$ is another critical exponent.  The way in
which dimensional analysis is preserved is that if we write out the
functional dependence more precisely, it is found that $\hat G(k, T_c)
\sim k^{-2} (k\ell)^\eta$.  This is a rather counter-intuitive result
because one would naively have expected that $\ell$ could be neglected
in the functional form of $G(k, T_c)$ since it is so much smaller than
the correlation length $\xi(T)$, which has in fact diverged at the
critical temperature.  The fact that $\hat G(k, T_c)$ retains a
dependence on $\ell$, i.e. is of the form $G(k, \ell, T_c)$, in what
would otherwise have been a purely self-similar regime is incomplete
similarity, or more descriptively, scale interference.  Complete
similarity then corresponds to the assumption in mean field theory,
namely that $\eta=0$, and correspondingly $\hat G(k, T_c)$ is a pure
power law with no dependence on $\ell$.

In turbulence, the scale interference is a statement about the inertial
regime behavior.  In the limit of $\Re \rightarrow \infty$ the inertial
regime is generally thought to describe the dissipationless transfer of
energy from one scale to another and the K41 assumption is that this is
independent of the large scale of forcing $L$ or the Kolmogorov scale
$\eta_K$ beyond which dissipation sets in.  These assumptions uniquely
determine the form of the energy spectrum $E(k) \sim \bar\epsilon^{2/3}
k^{-5/3}$ in the inertial range $2\pi/L \ll k \ll 2\pi/\eta_K$, and
obey complete similarity.  However, if the assumption of complete
similarity is not valid, then $E(k)$ can actually be a function of both
$k$ and $L$ (the inertial range is sensitive to the manner of turbulent
forcing) or $k$ and $\eta_K$ (the inertial range is sensitive to the
dissipative processes at small scales).  These considerations are
reflected in Kolmogorov's refined similarity hypothesis, which assumes
that the scaling of the longitudinal velocity difference
\begin{equation}
\delta
v_\ell^2 \equiv \langle ([\vec v (\vec r + \vec n \ell) - \vec v (\vec
r)]\cdot \vec n)^2 \rangle
\label{eqn:longvel}
\end{equation}
with an inertial range length scale $\ell$
varies as
\begin{equation}
\delta v_\ell^2 \sim \ell^{2/3 + \eta}.
\label{eqn:inter}
\end{equation}
In this expression, $\eta$ is the intermittency exponent, which arises
as an anomalous scaling exponent characterizing the average dissipation
over a neighbourhood whose dimension is $\ell$.

To see how incomplete similarity modifies the scaling law for the
friction factor, we present an argument due to Mehrafarin and Pourtolami
\cite{MEHR08}, that is in the spirit of Kadanoff's block spin
construction \cite{kadanoff1966}.  The friction factor is assumed to
depend on $\delta v_\ell$ and the mean flow speed in the pipe $U$
through a decomposition of the Reynolds stress
\begin{equation}
\tau_R \sim \rho \delta v_\ell U,
\label{eqntaudecomp}
\end{equation}
where $\rho$ is the fluid density.  The wall roughness $r$ sets the
scale for the transfer of momentum to the wall, so we set $\ell=r$, and
the friction factor becomes
\begin{equation}
f \sim \delta v_r/U.
\label{fric}
\end{equation}
Due to the assumed
anomalous scaling of the longitudinal velocity difference, we see that
the friction factor transforms under a scale transformation of the
roughness elements $r\rightarrow \lambda r$ as
\begin{equation}
f\sim r^\alpha \qquad \alpha \equiv \frac{1}{3} + \frac{\eta}{2}.
\label{eqn:ffscaling}
\end{equation}
The same scale transformation will affect the Reynolds number also,
because on the scale of the roughness, the viscosity will scale as
$r\delta v_r$, leading to the transformed Reynolds number
\begin{equation}
\Re \rightarrow  \lambda^{-(\frac{4}{3} + \frac{\eta}{2})} \Re \quad
\mbox{as }
r\rightarrow \lambda r
\label{eqn:Re-scaling}
\end{equation}
These results show that the friction factor is a
generalized homogeneous function, with
\begin{equation}
f \left(\lambda\frac{r}{D}, \lambda^{-(1+\alpha)} \Re
\right)= \lambda^\alpha f \left(\frac{r}{D}, \Re \right)
\label{eqn:homof}
\end{equation}
Setting the arbitrary scale
factor $\lambda \propto \Re^{1/(1+\alpha)}$ we obtain the generalization
of Eq. (\ref{eqn:data-collapse}) in the form
\begin{equation}
f \left(\frac{r}{D}, \Re \right) =  \Re^{-({2+3\eta})/({8 + 3\eta})} F\left(\frac{r}{D}
\Re^{{6}/({8+3\eta})} \right)
\label{eqn:gen-f}
\end{equation}
where the universal scaling function $F(z)$ behaves as $z^\alpha$ for
large $z$ and tends to a constant for small $z$.  A subtlety of this
derivation is that the Kolmogorov scale itself varies in a way that
depends on the intermittency exponent:
\begin{equation}
\eta_K \sim \Re^{-6/(8+3\eta)}
\label{eqn:Re-interm}
\end{equation}

In order to determine the exponent $\eta$, Nikuradze's data are plotted
analogously to \cite{GOLD06}, but generalized according to Eq.
(\ref{eqn:gen-f}), and the value $\eta$ adjusted to optimize the data
collapse.  The resulting value, $\eta = 0.02$ is consistent with
previous spectral estimates based on directly measuring the velocity
fluctuations and determining $E(k)$.  The result is rather remarkable:
eight years before Kolmogorov was to formulate the central scaling law
of the mean field theory of turbulence, Nikuradze had measured the
anomalous scaling exponent merely by accurate measurements of the
pressure drop along a turbulent pipe!

The argument presented above \cite{MEHR08} is truly in the spirit of
Kadanoff's block spin construction, because of Eq. (\ref{eqn:homof}).
Under a scale transformation, the friction factor as a function of its
two arguments retains {\it the same functional form}, but the arguments
get scaled in particular ways.  This is analogous to the way in which
Kadanoff derived a functional equation for the Helmholtz free energy
\cite{kadanoff1966}.  He assumed that under scale transformation, the
Hamiltonian of a spin system retained its functional form, but the spin
degrees of freedom were scaled in a particular way, to take into
account that microscopic spins transformed into block spins. Kadanoff's
assumption generates the homogeneous functional form of the free energy
per spin, and leads to Widom's scaling relations.  However, the
Kadanoff block spin picture is not capable of computing the actual
critical exponents.  The reason is that Kadanoff's assumption that the
functional form is invariant is wrong in general, because
coarse-graining introduces new non-local couplings in the effective
Hamiltonian governing the coarse-grained degrees of freedom.  Thus the
Hamiltonian changes during coarse-graining, and it was Wilson's great
achievement to recognize that this can be taken into account by writing
down recursion relations for the way in which the coupling constants
vary under coarse-graining.  Moreover, if these recursion relations
flow to a fixed point, then the functional form of the Hamiltonian is
invariant (by definition) at the renormalization group fixed point, and
in the neighbourhood of this fixed point, the critical exponents can be
obtained from a linearization of the coarse-graining transformation.

How could we in principle extend Goldenfeld, Mehrafarin and
Pourtolami's arguments to obtain a genuine RG calculation of the
anomalous dimensions of turbulence?  The answer must be that one can
emulate Wilson's argument for at least this wall-bounded turbulent
shear flow, by obtaining an approximate formula for the way in which
the friction factor changes under scale transformations of the
roughness, e.g. using decimation in one dimension along the pipe.
However, this program has not been completed to date, because it would
require a detailed calculation of the transformation of the friction
factor under coarse-graining.  In fact, even a calculation of the
exponents from Eq. (\ref{eqn:Blasius}) and Eq. (\ref{eqn:Strickler})
without anomalous dimensions is only possible using heuristic momentum
balance arguments due to Gioia and Chakraborty \cite{GIOIA06}, which we
briefly summarize.

\subsection{Spectral link and a fluctuation-dissipation
theorem for turbulence}

The starting point is the decomposition of the stress, Eq.
(\ref{eqntaudecomp}), where the length scale $\ell$ is either $r$ or
the Kolmogorov scale $\eta_K$, depending on whether wall roughness or
molecular viscosity makes the greater contribution to the dissipation.
Crudely we can represent this by writing
\begin{equation}
\ell = r + a \eta_K
\label{dissip}
\end{equation}
where $a$ is a constant of order unity.  In Eq. (\ref{fric}), we
estimate $\delta v_\ell$ by using the definition of the energy spectrum
as
\begin{equation}
E(k) \equiv \frac{d}{dk}\left(\frac{1}{2}\delta \hat v_k^2\right)
\label{spec}
\end{equation}
so that
\begin{equation}
\delta v_\ell = \sqrt{ \int^{\infty}_{2\pi/\ell} E(k)\,dk}
\label{vell}
\end{equation}
This leads to the remarkable formula
\be
f \propto \sqrt{ \int^{\infty}_{2\pi/\ell} E(k)\,dk}
\label{fdfric}
\ee
which explicitly connects the velocity fluctuations at small scales
with the large scale dissipation $f$.  In this sense, \eqnp{fdfric} is
a sort of fluctuation-dissipation theorem, establishing the explicit
sense in which we can say that turbulence can be usefully understood as
a non-equilibrium steady state.

If we use the K41 form
\be
E(k)\propto k^{-5/3}
\label{K41}
\ee
and the fact that, in the absence of intermittency, the Kolmogorov scale $\eta_K \sim
\Re^{-3/4}$ (from \eqnp{eqn:Re-interm}), we find that $f\sim (r/D)^{1/3}$ for large Re,
when the important dissipation scale is $\ell \sim r$.  This is just
the Strickler scaling, \eqnp{eqn:Strickler}.  On the other hand, at
smaller values of Re, but still in the turbulent regime, the dissipation from wall
roughness is insignificant compared to that arising from the cascade to
small molecular viscosity scales, and $\ell \sim \eta_K$.  This leads
to $ f\sim Re^{-1/4}$, nicely recovering \eqnp{eqn:Blasius}.

The predictions of this approach are experimentally testable, because
the friction factor scaling with Reynolds number (and wall-roughness)
is determined precisely by the energy spectrum.  So in this statistical
mechanical approach, the macroscopic dissipative behavior, as
quantified by the friction factor, reflects the nature of the turbulent
state through the energy spectrum functional form.  On the other hand,
the standard theory of wall-bounded turbulent shear flows is not able
to make such a connection.  The connection between the energy spectrum
and the macroscopic flow properties in steady-state turbulent flows has
been termed the spectral link
\cite{TRAN10,gioia2010spectral,kellay2012testing,zuniga2014spectral}.

The spectral link predictions have been tested experimentally in
two-dimensional soap films, where turbulent energy spectra with
exponents -5/3 and -3 can be created, consistent with an inverse energy
cascade and a forward enstrophy cascade respectively.  The spectral
link theory predicts that the friction factor, $f$, in the smooth-wall
Blasius regime should scale with Reynolds number, Re, as $f \sim
\Re^{-\alpha}$ with $\alpha=1/4$ and $1/2$ respectively for the inverse
energy and forward enstrophy cascades.  These results were indeed
obtained in careful experiments \cite{TRAN10}.  At the present time,
there are no direct experimental results to test the Strickler
exponents.  However, direct numerical simulations of the flow in two
dimensions with rough walls have been performed by using a conformal
map to transform the flow domain into a strip, and then using spectral
methods on the resulting transformed Navier-Stokes equations in a strip
\cite{GUTT09}.  These simulations demonstrated energy spectra
consistent with both the forward enstrophy cascade and the inverse
energy cascade, and the corresponding friction factor scalings were
consistent with the predictions of the momentum transfer theory, and
exhibited roughness-induced criticality up to $\Re=64,000$.

Sometimes it is objected that turbulence in two dimensions is not the
same as it is in three dimensions, because of the absence of vortex
stretching.  This is of course true, but irrelevant to the perspective
expressed here.  What is important is the presence of a cascade,
regardless of the precise mechanism from which it emerges.  The
statistical properties of the velocity fluctuations at small distances
compared to the integral scale determine the large-scale
dissipative processes in turbulence fluids, and the use of statistical
mechanical reasoning generates new conceptual insights, such
as the idea that friction factors should depend on the energy
spectrum, rather than a mean velocity profile as in the standard theory
of wall-bounded turbulent shear flows.  Statistical mechanics also
generates new experimental predictions, which have been partially
tested in turbulent soap films.

Despite these advances, there is a lot that needs to be done.  The
momentum transfer argument for the calculation of the friction factor
functional form is too simple, and in particular does not distinguish
between streamwise and transverse correlations.  The expression
\eqnp{eqntaudecomp} is a crude Reynolds decomposition of the
interaction between turbulence and the mean flow, and omits many
important details, some of which could perhaps now be addressed at
least in two dimensional flows \cite{falkovich2016interaction}.
The connection between the macroscopic dissipation and the microscopic
velocity fluctuations arises in a rather ad hoc fashion, and although
it expresses the same connection as described by the
fluctuation-dissipation theorem, a more formal analysis of the
connection would illuminate the way in which turbulence is a
non-equilibrium steady state, exhibiting fluctuation-dissipation
theorem properties in a way that makes contact with fluctuation
theorems far from equilibrium \cite{harada2005equality,prost2009generalized,seifert2012stochastic}.

\section{Statistical mechanics of the transition to turbulence}
\label{sec:transition2turbulence}

We turn now to the remarkable scaling behavior near the onset of pipe
turbulence around Reynolds number 2000
\cite{hof_lifetime,avila2011onset}, which is now convincingly
established as exhibiting the scaling behavior of a well-understood
non-equilibrium phase transition: directed percolation (DP).

Briefly, DP is a lattice model of a contact process, with a preferred
direction. In the variant known as bond directed percolation, with
probability $p$, bonds are placed on a diamond lattice oriented at 45
degrees to the preferred direction, and the resulting percolation
cluster is grown along the preferred direction, typically starting from
a single site, for example.  It is found that below a critical value
$p_c$, the cluster will eventually stop growing, so that there will not
be a path which percolates through the system. Above $p_c$, a
percolating path can be found, and the point $p=p_c$ exhibits scaling
behaviour similar to that found in equilibrium critical points.
However, the directed percolation cluster exhibits anisotropic scaling,
with different correlation lengths along and perpendicular to the
preferred direction.  These correlation lengths diverge at $p_c$, but
with different exponents $\nu_\bot$ and $\nu_\parallel$.  For a
thorough introduction and review of DP, see (e.g.)
\cite{hinrichsen2000}).

\subsection{Lifetime of turbulent puffs}
\label{sub:puff}

The transition to turbulence in pipes was of course originally studied
by Reynolds \cite{REYN83} but it would be 130 years before measurements
of the statistical behavior of the lifetime of turbulence could be
performed in a stable and systematic way.  Here we describe the recent
progress in a selective way to focus on the non-equilibrium statistical
physics aspects of the problem.  Other recent accounts summarize
additional aspects of the problem
\cite{cvitanovic2013recurrent,song2014deterministic,manneville2015transition,pomeau2016long,barkley2016}.

The breakthrough measurements in 2006 \cite{hof2006flt} revealed a
surprising and unanticipated result: the so-called mean lifetime $\tau$
of turbulence fluctuations (puffs) about an initially laminar flow
state increases rapidly as a function of Re, but there is no apparent
transition or vertical asymptote.  Indeed, it seemed that the data
could be represented to a good approximation by
\be
\ln \tau \propto \Re. \label{exptau}
\ee
These findings led its authors to speculate that the phenomenon of
turbulence was in fact just a very long-lived transient state, and that
there was neither a sequence of bifurcations \cite{landau1944problem}
nor a sharp transition between laminar and turbulent flows, as had been
previously believed.  This was especially surprising because for
several decades, it had been expected that the transition to turbulence
followed the pattern firmly established for the routes to chaos, in
particular classic work by Ruelle and Takens on strange attractors
\cite{ruelle1971nature} and Feigenbaum on period doubling
\cite{feigenbaum1978quantitative}.  These works on low-dimensional
dynamical systems have influenced recent attempts to describe
turbulence using the language and techniques of dynamical systems
theory
\cite{cvitanovic2013recurrent,song2014deterministic,barkley2016}.  This
has yielded many interesting insights into, and even experimental
measurements of, deterministic spatially-localized and unstable exact
solutions of the Navier-Stokes equations; but by its nature this
approach is less well-suited to explaining the statistical properties
of the transition which concern us here.

The early results suggesting that the mean lifetime scales as exp (Re)
were superseded in 2008 by a remarkable {\it tour de force}
\cite{hof_lifetime}, which established that the divergence is actually
a double exponential (super-exponential) function of Re:
\be
\ln \ln \tau \propto \Re,
\label{doubleexp}
\ee
with scaling observed over six decades in decay rate $1/\tau$.  The
functional form exp(exp(Re) was argued to arise in some way as a
low-dimensional chaotic supertransient
\cite{crutchfield1988art,tel2008cts} but the manner in which the system
size was replaced by the Re in these models, and precise details of how
this could arise and connect to the Navier-Stokes equations are
unclear.

\subsection{Extreme value statistics}
\label{sub:extreme}

An alternative approach to interpreting the super-exponential behavior
is based on the fact that the laminar state is an absorbing state, into
which patches of turbulence will decay in the aftermath of a large
enough spatial fluctuation in turbulent intensity
\cite{schneider2008lifetime,WK09}.  The decay rate of turbulence is
then proportional to the probability that the largest fluctuation
overcomes the Re-dependent threshold for the turbulence-laminar
transition \cite{nigel_evs}.  By virtue of involving the largest
fluctuation, this probability is calculated from the appropriate
extreme value distribution \cite{fisher_tippett,gumb58}.

Extreme value theory answers the following question.  Given a set of
independent, identically distributed random variables ${x_i}$
($i=1\dots N$), what is the probability distribution of the maximum
$X_N \equiv \textrm{max} \{x_i\}$?  Unlike the central limit theorem,
which provides the unique answer (in most circumstances) to the
question of what is the probability distribution of the mean of the
random variables (i.e. the normal distribution), the extreme value
theorem has three possible answers, depending on the asymptotics of the
probability density governing the original variables ${x_i}$.  For most
cases, where this density decays rapidly enough at infinity (as an
exponential or faster), the appropriate probability density is the Type
I Fisher-Tippett distribution, sometimes also known as the Gumbel
distribution \cite{gumbel1935valeurs}.  Its cumulative distribution has
the form:
\be
F(x) = \exp \left( -e^{-(x-\mu)/\beta} \right) \label{Gumcumul}
\ee
where $\mu$ and $\beta$ are parameters that set the location and scale
respectively of the distribution.  Using this distribution, and Taylor
expanding the probability distribution for the largest fluctuation in
Re (since the range of Re over which the transition occurs is small:
$1800 < \Re < 2000$), the super-exponential form \eqnp{doubleexp} is
recovered \cite{nigel_evs}.

At higher values of Re, the puffs were found not only to decay, but
also to split.  In the puff-splitting regime, the world-lines of puffs
trace out a complex branching pattern, observed in experiment and also
highly-resolved direct numerical simulations (DNS).  Both the decay
rate and the rate of splitting followed super-exponential scaling laws
with Re, the former increasing with Re and the latter decreasing.
Their crossover at $Re\approx 2040$ is interpreted as the single
distinguishing Reynolds number in the transitional region, and is
identified as the critical value $\Re_c$ \cite{avila2011onset}.

\subsection{DP and the decay of turbulent puffs}
\label{sub:decay}

A separate approach to the problem stems from Pomeau's prescient
intuition \cite{pomeau} (but see \cite{pomeau2016long} for a
counter-argument!) that the laminar state is an absorbing state, into
which patches of turbulence will decay in the aftermath of a large
enough spatial fluctuation in turbulent intensity
\cite{schneider2008lifetime,WK09}.  Including the diffusion of
turbulence, i.e. the spread of turbulent intensity into nearby laminar
regions, suggests that the laminar-turbulence transition is governed by
a contact process in the universality class of directed percolation
(DP) \cite{hinrichsen2000}.  This conclusion follows because DP is
widely believed to be the universality class for any local
non-equilibrium absorbing process
\cite{janssen1981,grassberger1982phase}.

Pomeau's initial suggestion \cite{pomeau} was followed up by
simulations of the damped Kuramoto-Sivashinsky equation, where
spatiotemporal intermittency coexists with locally uniform domains in a
way that seems reminiscent of DP; in particular, as a control parameter
is varied, the equation's order parameter evolves as a continuous
process beyond a threshold where it jumps discontinuously through a
sub-critical bifurcation \cite{chate1987transition}.  A much later
study \cite{barkley2011simplifying}, motivated by a perceptive analogy with excitable media, used a
model 1 + 1 dimensional nonlinear partial differential equation coupled
to a tent map. Numerical simulations showed a similar phase diagram to
the experiments in pipe flow turbulence, with laminar, metastable
turbulence and spatiotemporal intermittency as the control parameter
analogous to Re was increased.  Furthermore, the order parameter in the
spatiotemporal intermittent phase scaled in a way consistent with the
order parameter scaling of DP.

To test the DP scenario in more quantitative detail, and to give an
interpretation to the super-exponential behavior, simulations of DP
were performed in the geometry of a pipe and emulating the conditions
of the experiments \cite{sipos2011directed}.  The basic idea is that an
occupied site on a lattice corresponds to a turbulent correlation
volume, and an empty site corresponds to a laminar region.  Starting
from a localized puff of \lq\lq turbulence" and for $p < p_c$, the DP
region eventually dies away, whereas for $p>p_c$ it spreads and fills
the pipe. The mean lifetime $\tau$ of puffs could thus be measured,
following the procedure used in experiment \cite{hof_lifetime}.  These
numerical experiments recapitulated the super-exponential behavior
observed in the experiments, and moreover provided an alternative
rationale for the super-exponential distribution for the decay rate.
In DP, the history of occupied sites in successive time slices traces
out a complex network of paths, which ends when the last turbulent or
occupied site is reached.  Thus the lifetime of turbulence is the
length of the {\it longest} path in the DP simulation, and its
probability density would follow extreme value statistics
\cite{sipos2011directed,bazant}.

These arguments can be extended to the case where there is
puff-splitting (i.e. for $p > p_c$), and simulation results confirm the
predicted super-exponential dependence for the splitting rate as well
\cite{dpturb2016}.  It may seem surprising that DP itself exhibits a
super-exponential scaling law.  One might wonder why the timescales do
not diverge at a the percolation threshold $p_c$ with the appropriate
power-law divergence.  The answer turns out to be subtle and related to
the precise way in which $\tau$ for decay and splitting is measured
\cite{dpturb2016}.  In fact, it is possible in principle to extract the
expected power-law divergences from experimental data \cite{dpturb2016}
even while they appear to show super-exponential behavior and thus no
signature of a critical point.

To summarise: experimental data and theory strongly suggest that the
laminar-turbulence transition in pipes is in the universality class of
directed percolation.  Recent experiments on ultra-narrow gap large
aspect ratio Couette flow \cite{lemoult2016directed} and on channel
flow \cite{sano2016universal} report measurements of the critical
exponents and in the case of the Couette flow, even the universal
scaling functions.

\subsection{Landau theory for laminar-turbulence transition}
\label{sub:landau}

How is it possible that a driven fluid flow in a spatial continuum
could behave precisely like a discrete lattice model from
non-equilibrium statistical mechanics, surely an approximation at best?
Such exactitude is unprecedented in fluid mechanics but the underlying
explanation rests with the theory of phase transitions
\cite{goldenfeld1992lectures}, of which it is our contention that the
laminar-turbulence transition is an example. There it is
well-established that universal aspects of phase transitions, such as
the phase diagram, critical exponents and scaling functions are all
described {\it exactly\/} by an effective coarse-grained theory (\lq\lq
Landau theory\rq\rq) that contains only the symmetry-allowed collective
and long-wavelength modes, without requiring excessive realism at the
microscopic level of description. Being based on symmetry principles,
the individual symmetry-allowed terms in Landau theory do not require
detailed derivation from the microscopic level of description.  This is
fortunate given that there is usually no good, uniformly valid
approximation scheme to derive formally and systematically these terms
and their coefficients from first principles.  This is true in
equilibrium phase transitions, and all the more so in the
laminar-turbulence transition, which occurs far from equilibrium.

In fact, it is neither necessary nor desirable to derive the
coarse-grained effective theory from a microscopic description, because
any such derivation would need a small parameter and would thus only
have limited validity due to the analytical approximations made. A
familiar example of this situation is that even though the
Navier-Stokes equations can be derived from Boltzmann's kinetic
equations for gases, such a derivation would imply that the
Navier-Stokes equation is only valid for dilute gases.  In fact, the
Navier-Stokes equations are an excellent description for dense liquids
as well, and can be obtained by perfectly satisfactory phenomenological
and symmetry arguments.  The derivation from Boltzmann's kinetic theory
is inherently limited by the regime of validity of the kinetic
theory---low density---and this leads to an unnecessarily restrictive
derivation of the equations of fluid dynamics.  Returning to phase
transitions, the reason why an analytical derivation of the
coarse-grained theory is unnecessary is that even if the coefficients
of the terms could be computed in the order parameter expansion of the
Landau theory, they do not come into the exponents or scaling functions
anyway, and thus they do not affect the critical behavior.  In the case
of the transition to turbulence, the strategy then is to construct an
effective theory that is valid near the transition.  This effective
theory would be an exact representation of the critical behavior of the
laminar-turbulent transition, and as is often the case, could
potentially be mapped into one of the canonical representatives of a
known universality class.  That universality class turns out to be DP.

In order to build an effective theory for the transitional turbulence
problem by constructing the symmetry-allowed collective and
long-wavelength modes, the analytical difficulties are acute.
Therefore, to avoid approximations which are difficult to justify
systematically, direct numerical simulation was used to identify the
important collective modes which exhibit an interplay between
large-scale fluctuations and small-scale dynamics at the onset of
turbulence, and thence to write down the corresponding minimal
stochastic model, in the spirit of the Landau theory of phase
transitions \cite{shih2016ecological}.

\begin{figure}
\begin{center}
\includegraphics[width=\textwidth]{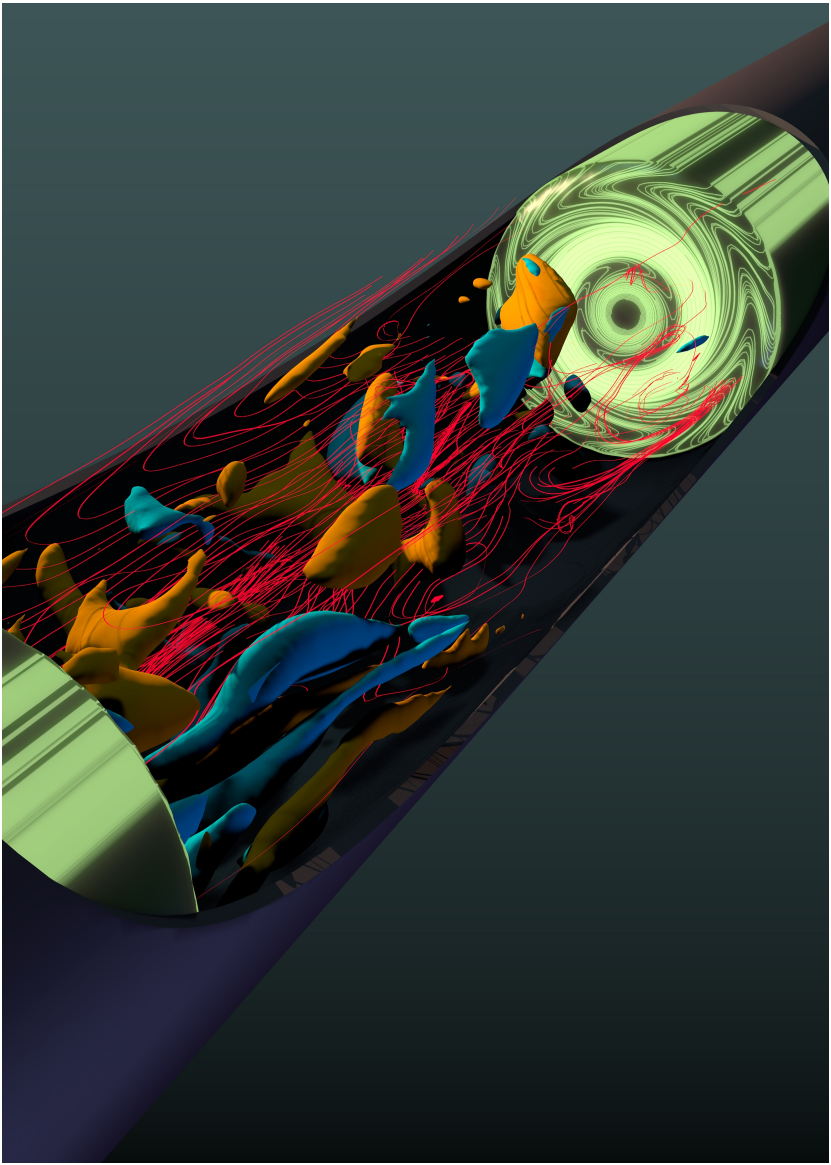}
\end{center}
\caption{Cutaway view of DNS for pipe flow near the
transition to turbulence, showing the zonal flow (green), isosurfaces
of Reynolds stress (blue and orange) and streamlines (red).}
\label{cutaway}
\end{figure}

\subsection{Zonal flow and predator-prey dynamics}
\label{sub:zonal}

The result of the numerical simulations was the identification of a
collective mode near the transition which regulates turbulence but is
itself generated by the turbulence \cite{shih2016ecological}.
Technically speaking this mode is a zonal flow.  It is purely
azimuthal, but has radial and time dependence, and no dependence on
axial coordinate $z$.  The mode is not driven by the pressure drop
along the axis of the pipe, in contrast to the mean flow.  Instead it
is driven by the turbulence itself, in particular arising from the
anisotropy of the Reynolds stress tensor. The mode shears the
turbulence and thus has the effect of reducing the anisotropy of the
turbulent fluctuations.  In turn, this reduces the intensity of the
zonal flow.  Once the zonal flow intensity has diminished, the
turbulence is no longer sheared so strongly and so is less suppressed
than before.  As a result the energy in the turbulent modes increases,
and the cycle begins again.  This narrative of the interplay between
zonal flow and turbulence was established by measuring the energy in
the zonal flow and turbulent degrees of freedom, the azimuthal flow
velocity and the Reynolds stress \cite{shih2016ecological}.

In general, it is the case that zonal flows are driven by statistical
anisotropy in turbulence, but are themselves an isotropizing influence
on the turbulence through their coupling to the Reynolds
stress \cite{sivashinsky1985negative,bardoczi2014experimental,parker2014generation}.  The interplay between
zonal flow suppression of turbulence and turbulence initiation of zonal
flow has also been reported in thermal convection in a variety of
geometries \cite{goluskin2014convectively,von2015generation}.
Originally predator-prey behavior was proposed by Diamond and
collaborators \cite{diamond1994,kim2003,itoh2006} many years ago in the
context of the interaction between drift-wave turbulence and zonal
flows in tokomaks. The predator-prey oscillations were recently
observed in tokomaks \cite{estrada2010,conway2011,xu2011,estradaPRL2011,schmitz2012} and in
a table-top electroconvection analogue of the L-H transition
\cite{bardoczi2014experimental}.

\subsection{Lotka-Volterra equations in transitional turbulence}
\label{sub:lotka}

The activation-inhibition nature of the interplay described here
parallels that which occurs in predator-prey ecosystems.  Predator-prey
ecosystems exhibit the following well-known behavior.  A prey acts as a
source of food for a predator, and thus the predator population rises.
However, under increased predation, the prey population begins to
decline.  As a result the predator population subsequently declines as
well.  With reduced predation pressure, the prey population begins to
rise, and the cycle begins again.  This behaviour is typically modeled
by the Lotka-Volterra equations
\cite{lotka1910contribution,volterra1927variazioni,renshaw1993modelling}
for the population of predator $A$ and prey $B$:
\begin{eqnarray}
\dot A &&= pAB -dA\\
\dot B &&= bB(1-B/\kappa) -pAB
\label{lotka}
\end{eqnarray}
where time derivative is denoted by a dot, $p$ is predation rate, $d$
is predator death rate, $b$ is prey birth rate and $\kappa$ is the
carrying capacity (i.e. the maximum amount of prey that the ecosystem
nutrient supply can support).

An outline of how to derive the predator-prey equations in pipe
transitional turbulence is as follows, modeled after efforts to obtain
such equations heuristically in tokamak physics \cite{diamond1994}. We
start by sketching the form of an equation describing the time
variation of the energy of turbulent modes, due to local instabilities
and the likely interaction with the zonal flow.  The basic premise is
that there is a primary instability generating turbulence at small
scales, probably arising from the interaction of localized unstable
modes such as periodic orbits.  The energy of turbulent fluctuations
$E$ at the relevant wavenumber or range of wavenumbers will have three
main contributions to $dE/dt$. The first is the primary linear
instability of the form $\propto E$. The second term will be of higher
order, describing eddy interactions through some sort of non-local
scattering kernel or triad processes. We will make the usual ansatz
that near a phase transition, it is permissible to replace the
non-local kernel by local terms describing eddy damping, of the form
$\propto - E^2$. Finally, from the numerics we know that turbulent
fluctuations are suppressed by interaction with the zonal flow.  What
should be the form of this interaction? The zonal flow is a collective
shear mode; the azimuthal velocity component $u_\theta$
experiences shear in the radial direction $r$, and will be denoted by
$\Omega \equiv \partial \langle \overline{u}_\theta(r)\rangle/\partial
r$.  Here $\theta$ is the azimuthal direction, and
$\overline{u}_\theta(r)$ represents the purely azimuthal component of
the zonal flow that is spatially uniform in the longitudinal direction,
indicating it it not driven by pressure gradients in pipe flows.  The
damping should occur through interaction between the Reynolds stress
and $\Omega$, but should be independent of the direction of the shear.
Thus, the most generic coupling between the turbulence and $\Omega$
should be proportional to both $E$ and $U\equiv \Omega^2$.

These considerations suggest that
\begin{equation}
\frac{dE}{dt} = \gamma_0 E - \alpha_1E^2 - \alpha_2 EU
\label{dedt}
\end{equation}
where $\gamma_0$, $\alpha_1$ and $\alpha_2$ are constants.

Next we sketch an outline of how one can obtain a description of the
zonal flow equation of motion. The starting point in the Reynolds
momentum balance equation, which we will write in the approximate form
for the zonal flow velocity field:
\begin{equation}
\frac{\partial\vtr}{\partial t} = -\frac{\partial\langle
\tilde{v_r}\tilde{v_\theta}\rangle}{\partial r} - \mu \langle\vtr\rangle
\label{zf}
\end{equation}
where $\mu$ is some damping coefficient, the tilde denotes
fluctuation component and $\langle
\tilde{v_r}\tilde{v_\theta}\rangle$ is the Reynolds stress in the azimuthal direction.  In Eq. \ref{zf}, we have omitted  terms that
are in principle present from the Reynolds equation, but either vanish
due to the azimuthal average or are small compared to the terms
retained. We do not have a fully systematic derivation of this
equation.  However, we have measured the right and left hand terms of
this equation in the DNS and the results show that these terms do
indeed track one another \cite{shih2016ecological}.  Taking the radial
derivative of Eq. \ref{zf} leads to
\begin{equation}
\partial_t\Omega = -\partial_r^2\langle
\tilde{v_r}\tilde{v_\theta}\rangle - \mu\Omega
\label{rseom}
\end{equation}

This equation is not closed of course, but we can make progress with
scaling arguments. We conjecture that $\langle
\tilde{v_r}\tilde{v_\theta}\rangle$ is quadratic in velocity
fluctuations and therefore should be proportional to $E$.  Note that
the term $\langle \tilde{v_r}\tilde{v_\theta}\rangle$ vanishes by
symmetry in an isotropic flow, but is non-zero when the turbulent
fluctuations are anisotropic and coupled to the zonal flow which
provides a local directionality to the velocity fluctuations.  This
suggests that $-\partial_r^2\langle \tilde{v_r}\tilde{v_\theta}\rangle
\propto +\Omega + O(\Omega^2)$ where it is important to note that the +
sign means that the Reynolds stress anisotropy is exciting the zonal
flow, and not damping it, corresponding to the DNS results. Putting
these scaling arguments together and multiplying through by $\Omega$
suggests that
\begin{equation}
\partial_t U = \alpha_3 EU - 2\mu U
\label{ome}
\end{equation}
where $\alpha_3$ is another phenomenological constant.  The equation
argued for above is basically a scalar equation, but a full
understanding of the interaction of mean flows or zonal flows with
turbulence anisotropy requires a detailed consideration of the full
tensor Reynolds equation, the spatial variation of the eigenvectors of
the stress tensor etc.  The heuristic derivation described here can be
checked by direct numerical computations, in principle, and we hope to
do this in the future.  The immediate consequence of \eqnp{dedt} and
\eqnp{ome} is that they have the form of the mean field Lotka-Volterra
equations \eqnp{lotka}, and thus would be expected to predict (at the
mean field level) the existence of predator-prey oscillations.

Unfortunately, the mean field Lotka-Volterra equations do not predict
population oscillations at all!  It is straightforward to see that for
finite $\kappa$, their steady state is a constant solution, and
certainly not a limit cycle. When $\kappa=\infty$, the equations have
oscillatory solutions but they are centers, not asymptotically stable
limit cycles, and the amplitude and phase depend on the initial
conditions.  Faced with the paradox posed by the apparent discrepancy
between the mathematical solution and the expectation based on the
verbal description of predator-prey populations, the standard
resolution is to invoke other biological factors such as predator
satiation and other forms of what is known as \lq\lq functional
response".  These effects lead to modifications of the Lotka-Volterra
equations, introducing nonlinearities that guarantee limit cycle
behavior.

The most satisfying resolution of the paradox, however, is that no new
biological factors need to be introduced at all: the non-oscillatory
prediction from the Lotka-Volterra equations is an artifact of the mean
field approximation.  As shown by McKane and Newman
\cite{mckane2005predator}, an individual-level model of predator-prey,
wherein each organism's birth, death and predatory activity is
simulated, leads to persistent population oscillations. Stochasticity
of individual birth, death, predation processes leads to multiplicative
noise in the effective equations at the population level.  These
equations can be calculated accurately using van Kampen's system size
expansion \cite{kampen1961power,van2011stochastic}.  In the limit of
infinite population size, the oscillations vanish, but for finite
system size, the oscillations experience a resonance effect which
amplifies them.  In spatially-extended systems, stochasticity locally
drives instabilities, leading to fluctuation-induced
Turing patterns or traveling waves if there are appropriate sources of nonlinearity
\cite{butler2009robust,biancalani2010stochastic}.

\subsection{Stochastic predator-prey model for transitional turbulence}
\label{sub:stoch}

\begin{figure}
\begin{center}
\includegraphics[width=\textwidth]{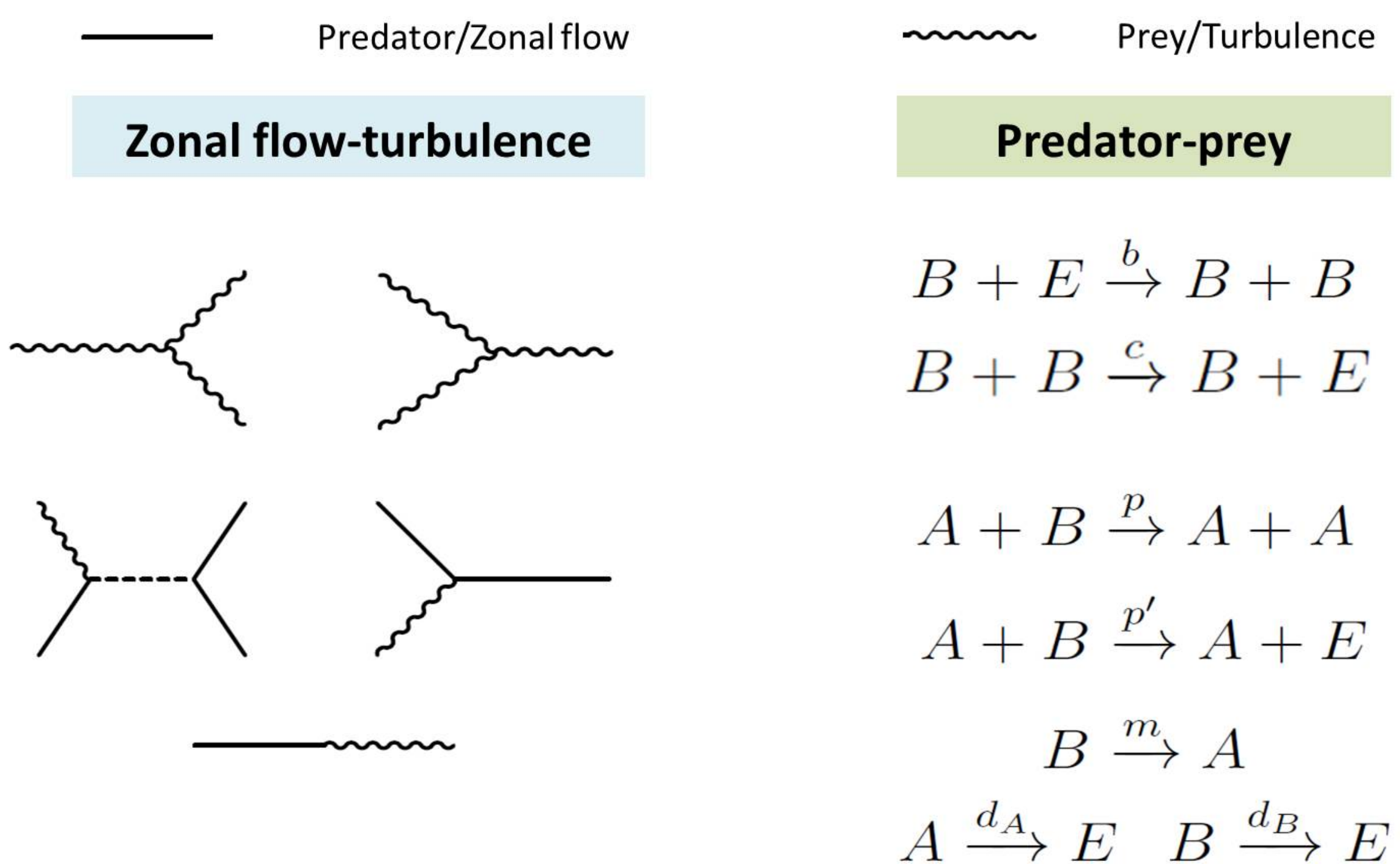}
\end{center}
\caption{Effective theory for interactions between small-scale
turbulence and large-scale zonal flow.  The allowed interactions
between turbulence modes (prey $B$, wiggly lines) and zonal flow
(predator $A$, straight lines) are shown, together with their
interpretation as processes describing birth, death and predatory
activity.  An allowed process at the lowest order corresponds to the
conversion between prey and predator with rate $m$, something that does not have a
direct realization in most biological systems.  The symbol $E$ denotes
the third trophic level in the ecosystem of turbulence, namely the food
that sustains the prey, which in the fluid dynamics picture is simply
the laminar flow state itself.  The symbols above the arrows denote
rate constants.  The definition of predator-prey processes is described in the text.  In the right column the predation process with rate $p'$ is not included in our model since it only renormalizes the predation coefficient in the prey equation in Eq. (\ref{lotka}).} \label{turb-zf}
\end{figure}

These considerations are significant for the interactions between
turbulence and zonal flows, as stochasticity, in effect, arises due to the
microscopic interactions between localized unstable modes of the fluid
flow.  The predator-prey nature of the interactions shows that
turbulence is the prey, whereas the zonal flow is the predator.  In
order to construct a Landau theory for the laminar-turbulence
transition, it is necessary to write down all possible interactions
between the turbulence and the zonal flow.  These are summarized in
Fig. (\ref{turb-zf}), and displayed in the language of stochastic
predator-prey processes.

The equations for the emergence of a zonal flow collective mode
interacting via activator-inhibitor/predator-prey kinetics with the
small-scale turbulence can be written down as a set of
spatially-extended  rate equations for the number of predator $A$ and
prey $B$ and nutrient (laminar flow) sites ($E$) as follows:

\begin{eqnarray}
&&A_i\xrightarrow{d_\text{A}}E_i,\qquad
B_i\xrightarrow{d_\text{B}}E_i,\qquad
A_i+B_j\xrightarrow[{\langle ij\rangle}]{p}A_i+A_j,\nonumber\\
&&B_i+E_j\xrightarrow[{\langle ij\rangle}]{b}B_i+B_j,\qquad
B_i\xrightarrow{m}A_i,\nonumber\\
&&A_i+E_j\xrightarrow[{\langle ij\rangle}]{D}E_i+A_j,\;\;\;
B_i+E_j\xrightarrow[{\langle ij\rangle}]{D}E_i+B_j.
\label{eqn:reactions}
\end{eqnarray}
where $d_\text{A}$ and $d_\text{B}$ are the death rates of A and B, $p$ is the
predation rate, $b$ is the prey birth rate due to consumption of
nutrient, $\langle ij\rangle$ denotes hopping to nearest neighbor
sites, $D$ is the nearest-neighbor hopping rate, and $m$ is the point
mutation rate from prey to predator, which models the induction of
the zonal flow from the turbulence degrees of freedom.

Remarkably, simulations of this predator-prey model, in a 2D strip
intended to represent the 3D pipe geometry of the original turbulence
experiments, reproduce the main features of the laminar-turbulence
transition. In this case, the control parameter turns out to be the
birth rate $b$ of the prey, and this is the analogue of Re
\cite{shih2016ecological}. First of all the phase diagram is reproduced
as a function of the birth rate $b$ of the prey, which plays the role
of Re. In particular there is a phase where no prey survive; then at
higher $b$, a phase where the prey and predator co-exist, but localised
regions of prey decay; then at still higher values of $b$, a region
where localized regions of prey split, so that the dynamics exhibits
the strong intermittency in space and time seen in the turbulence
simulations and experiments.  Furthermore, it is found that there is a
super-exponential variation of decay and splitting lifetimes on the
prey lifetime $b$ \cite{shih2016ecological}.

In addition to recapitulating the phenomenology of the
laminar-turbulence transition in pipes, the stochastic predator-prey
model \eqnp{eqn:reactions} can be mapped exactly into Reggeon field
theory \cite{mobilia2007,tauber2012} and this field theory itself has
long been known to be in the DP universality class
\cite{cardy1980,janssen1981}.  The connection between the
super-exponential scaling of timescales with Re and the expected
divergence at a critical value of Re is not explained in the original
work \cite{shih2016ecological}.  In fact, subsequently it has been
understood how to extract the dynamic critical exponents and the
divergence of lifetimes from the turbulence data, at least in
principle \cite{dpturb2016}.


\subsection{Summary}
\label{sub:summ}

In summary, we used DNS to identify the important
collective modes at the onset of turbulence---the predator-prey
modes---and then wrote down the simplest minimal stochastic model to
account for these observations. This model {\it predicts} without using
the Navier-Stokes equations the puff lifetime and splitting behavior
observed in experiment.  This approach is a precise parallel to that
used in the conventional theory of phase transitions, where one builds
a Landau theory, a coarse-grained (or effective) theory, using symmetry
principles. This intermediate level description can then be used as a
starting point for renormalization group analysis to compute the
critical behavior.  In this case, however, the statistical description
arises from non-equilibrium statistical mechanics, as the
predator-prey equations do not obey detailed balance.

Directed percolation arises due to the appearance of collective modes
near criticality whose fluctuations exhibit the characteristics of
stochastic predator-prey dynamics near the collapse of an ecosystem
from its coexistent state.  Both turbulence and predator-prey ecosystem
criticality reflect scaling laws that ultimately derive from extreme
value statistics, thus, establishing an unprecedented connection
between the laminar-turbulence transition, predator-prey extinction,
directed percolation, and extreme value theory.

Our approach is thus a precise parallel to the way in which phase
transitions are understood in condensed matter physics, and shows that
concepts of universality and effective theories are applicable to the
laminar-turbulence transition.

\section{Conclusion}

In this article, we have shown with concrete examples how turbulence
can usefully be viewed through the lens of non-equilibrium statistical
mechanics.  In particular we have shown how macroscopic dissipative
properties of wall-bounded turbulent shear flows, but especially pipe
flow, can be described using concepts from scaling theory for $\Re >
2000$. Moreover precise results from the conceptual framework of
renormalization group theory were used to identify the universality
class of the laminar-turbulence transition.  These results are
suggestive of a more profound connection between turbulent flows and
non-equilibrium statistical mechanics.

\begin{acknowledgements}
NG wishes to express his gratitude  to Leo P. Kadanoff for his
scientific inspiration, support, collaboration and friendship over many
decades. NG also wishes to thank P. Chakraborty, G. Gioia, W. Goldburg,
T. Tran, H. Kellay and N. Guttenberg for collaboration on the topics in
section \ref{sec:ff}.  We thank T.-L. Hsieh for collaboration on the
topics in section \ref{sec:transition2turbulence}.  We acknowledge
helpful discussions with L.P. Kadanoff, B. Hof, J. Wesfreid, P.
Manneville, D. Barkley and Y. Pomeau.  We thank N. Guttenberg for
technical assistance with figure (\ref{cutaway}).   This work was
supported in part by the National Science Foundation through grant
NSF-DMR-1044901.

\end{acknowledgements}



\bibliographystyle{spmpsci}
\bibliography{turbulence-transition-friction-factor-Kadanoff}

\begin{thebibliography}{10}
\providecommand{\url}[1]{{#1}}
\providecommand{\urlprefix}{URL }
\expandafter\ifx\csname urlstyle\endcsname\relax
  \providecommand{\doi}[1]{DOI~\discretionary{}{}{}#1}\else
  \providecommand{\doi}{DOI~\discretionary{}{}{}\begingroup
  \urlstyle{rm}\Url}\fi

\bibitem{avila2011onset}
Avila, K., Moxey, D., de~Lozar, A., Avila, M., Barkley, D., Hof, B.: The onset
  of turbulence in pipe flow.
\newblock Science \textbf{333}(6039), 192--196 (2011)

\bibitem{bardoczi2014experimental}
Bard{\'o}czi, L., Bencze, A., Berta, M., Schmitz, L.: Experimental confirmation
  of self-regulating turbulence paradigm in two-dimensional spectral
  condensation.
\newblock Physical Review E \textbf{90}(6), 063,103 (2014)

\bibitem{barenblatt1972self}
Barenblatt, G., Zel'Dovich, Y.B.: Self-similar solutions as intermediate
  asymptotics.
\newblock Annual Review of Fluid Mechanics \textbf{4}(1), 285--312 (1972)

\bibitem{barenblatt1996scaling}
Barenblatt, G.I.: Scaling, self-similarity, and intermediate asymptotics.
\newblock Scaling, Self-similarity, and Intermediate Asymptotics, by Grigory
  Isaakovich Barenblatt, pp. 408. ISBN 0521435226. Cambridge, UK: Cambridge
  University Press, December 1996. p. 408 (1996)

\bibitem{barkley2011simplifying}
Barkley, D.: Simplifying the complexity of pipe flow.
\newblock Physical Review E \textbf{84}(1), 016,309 (2011)

\bibitem{barkley2016}
Barkley, D.: Theoretical perspective on the route to turbulence in a pipe.
\newblock Journal of Fluid Mechanics \textbf{803}, P1 (2016).
\newblock \doi{10.1017/jfm.2016.465}.
\newblock \urlprefix\url{http://dx.doi.org/10.1017/S002211200600454X}

\bibitem{bazant}
Bazant, M.Z.: Largest cluster in subcritical percolation.
\newblock Phys. Rev. E \textbf{62}(2), 1660--1669 (2000).
\newblock \doi{10.1103/PhysRevE.62.1660}

\bibitem{biancalani2010stochastic}
Biancalani, T., Fanelli, D., Di~Patti, F.: Stochastic {T}uring patterns in the
  {B}russelator model.
\newblock Phys. Rev. E \textbf{81}(4), 046,215 (2010)

\bibitem{blasius1913ahnlichkeitsgesetz}
Blasius, H.: {Das {\"A}hnlichkeitsgesetz bei Reibungsvorg{\"a}ngen in
  Fl{\"u}ssigkeiten}.
\newblock Forschg. Arb. Ing.-Wes \textbf{134} (1913)

\bibitem{butler2009robust}
Butler, T., Goldenfeld, N.: Robust ecological pattern formation induced by
  demographic noise.
\newblock Phys. Rev. E \textbf{80}(3), 030,902 (2009)

\bibitem{cao1993conceptual}
Cao, T.Y., Schweber, S.S.: The conceptual foundations and the philosophical
  aspects of renormalization theory.
\newblock Synthese \textbf{97}(1), 33--108 (1993)

\bibitem{cardy2008non}
Cardy, J., Falkovich, G., Gawedzki, K.: Non-equilibrium statistical mechanics
  and turbulence.
\newblock Cambridge University Press (2008)

\bibitem{cardy1980}
Cardy, J.L., Sugar, R.L.: Directed percolation and {R}eggeon field theory.
\newblock Journal of Physics A: Mathematical and General \textbf{13}(12),
  L423--L427 (1980)

\bibitem{chate1987transition}
Chate, H., Manneville, P.: Transition to turbulence via spatiotemporal
  intermittency.
\newblock Physical Review Letters \textbf{58}, 112--115 (1987)

\bibitem{conway2011}
Conway, G.D., Angioni, C., Ryter, F., Sauter, P., Vicente, J.: Mean and
  oscillating plasma flows and turbulence interactions across the {L}-{H}
  confinement transition.
\newblock Phys. Rev. Lett. \textbf{106}, 065,001 (4 pages) (2011).
\newblock \doi{10.1103/PhysRevLett.106.065001}

\bibitem{crutchfield1988art}
Crutchfield, J., Kaneko, K.: {Are attractors relevant to turbulence?}
\newblock Physical Review Letters \textbf{60}, 2715--2718 (1988)

\bibitem{cvitanovic2013recurrent}
Cvitanovi{\'c}, P.: Recurrent flows: the clockwork behind turbulence.
\newblock Journal of Fluid Mechanics \textbf{726}, 1--4 (2013)

\bibitem{diamond1994}
Diamond, P.H., Liang, Y.M., Carreras, B.A., Terry, P.W.: Self--regulating shear
  flow turbulence: A paradigm for the {L}-{H} transition.
\newblock Phys. Rev. Lett. \textbf{72}, 2565--2568 (1994).
\newblock \doi{10.1103/PhysRevLett.72.2565}

\bibitem{edwards1964statistical}
Edwards, S.: The statistical dynamics of homogeneous turbulence.
\newblock Journal of Fluid Mechanics \textbf{18}(02), 239--273 (1964)

\bibitem{estrada2010}
Estrada, T., Happel, T., Hidalgo, C., Ascasíbar, E., Blanco, E.: Experimental
  observation of coupling between turbulence and sheared flows during {L}-{H}
  transitions in a toroidal plasma.
\newblock EPL \textbf{92}(3), 35,001 (6 pages) (2010)

\bibitem{estradaPRL2011}
Estrada, T., Hidalgo, C., Happel, T., Diamond, P.H.: Spatiotemporal structure
  of the interaction between turbulence and flows at the {L}-{H} transition in
  a toroidal plasma.
\newblock Phys. Rev. Lett. \textbf{107}, 245,004 (5 pages) (2011).
\newblock \doi{10.1103/PhysRevLett.107.245004}

\bibitem{EYIN94}
Eyink, G., Goldenfeld, N.: Analogies between scaling in turbulence, field
  theory and critical phenomena.
\newblock Phys. Rev. E \textbf{50}, 4679--4683 (1994)

\bibitem{falkovich2016interaction}
Falkovich, G.: Interaction between mean flow and turbulence in two dimensions.
\newblock Proc. R. Soc. A \textbf{472}, 20160,287 (2016)

\bibitem{feigenbaum1978quantitative}
Feigenbaum, M.J.: Quantitative universality for a class of nonlinear
  transformations.
\newblock Journal of statistical physics \textbf{19}(1), 25--52 (1978)

\bibitem{fisher1998renormalization}
Fisher, M.E.: Renormalization group theory: Its basis and formulation in
  statistical physics.
\newblock Reviews of Modern Physics \textbf{70}, 653--681 (1998)

\bibitem{fisher_tippett}
Fisher, R.A., Tippett, L.H.C.: Limiting forms of the frequency distribution of
  the largest or smallest member of a sample.
\newblock Proc. Cambridge Phil. Soc. \textbf{24}, 180--190 (1928)

\bibitem{GIOIA06}
Gioia, G., Chakraborty, P.: Turbulent friction in rough pipes and the energy
  spectrum of the phenomenological theory.
\newblock Phys.\ Rev.\ Lett. \textbf{96}, 044,502 (2006)

\bibitem{gioia2010spectral}
Gioia, G., Guttenberg, N., Goldenfeld, N., Chakraborty, P.: Spectral theory of
  the turbulent mean-velocity profile.
\newblock Physical Review Letters \textbf{105}(18), 184,501 (2010)

\bibitem{goldenfeld1992lectures}
Goldenfeld, N.: Lectures On Phase Transitions And The Renormalization Group.
\newblock Addison-Wesley Reading, MA (1992)

\bibitem{GOLD06}
{Goldenfeld}, N.: {Roughness-Induced Critical Phenomena in a Turbulent Flow}.
\newblock Phys. Rev. Lett. \textbf{96}, 044,503 (2006)

\bibitem{nigel_evs}
Goldenfeld, N., Guttenberg, N., Gioia, G.: Extreme fluctuations and the finite
  lifetime of the turbulent state.
\newblock Phys. Rev. E \textbf{81}(3), 035,304 (2010).
\newblock \doi{10.1103/PhysRevE.81.035304}

\bibitem{goluskin2014convectively}
Goluskin, D., Johnston, H., Flierl, G.R., Spiegel, E.A.: Convectively driven
  shear and decreased heat flux.
\newblock Journal of Fluid Mechanics \textbf{759}, 360--385 (2014)

\bibitem{grassberger1982phase}
Grassberger, P.: On phase transitions in {S}chl{\"o}gl's second model.
\newblock Zeitschrift f{\"u}r Physik B Condensed Matter \textbf{47}(4),
  365--374 (1982)

\bibitem{gumbel1935valeurs}
Gumbel, E.: Les valeurs extr{\^e}mes des distributions statistiques.
\newblock Annales de l'institut Henri Poincar{\'e} \textbf{5}(2), 115--158
  (1935)

\bibitem{gumb58}
Gumbel, E.: {Statistics of Extremes}.
\newblock Columbia University Press, New York, NY (1958)

\bibitem{GUTT09}
Guttenberg, N., Goldenfeld, N.: {Friction factor of two-dimensional
  rough-boundary turbulent soap film flows}.
\newblock Phys.\ Rev.\ E \textbf{79}(6), 65,306 (2009)

\bibitem{harada2005equality}
Harada, T., Sasa, S.I.: Equality connecting energy dissipation with a violation
  of the fluctuation-response relation.
\newblock Physical Review Letters \textbf{95}(13) (2005)

\bibitem{von2015generation}
von Hardenberg, J., Goluskin, D., Provenzale, A., Spiegel, E.: Generation of
  large-scale winds in horizontally anisotropic convection.
\newblock Physical review letters \textbf{115}(13), 134,501 (2015)

\bibitem{hinrichsen2000}
Hinrichsen, H.: Non-equilibrium critical phenomena and phase transitions into
  absorbing states.
\newblock Advances in Physics \textbf{49}(7), 815--958 (2000).
\newblock \doi{10.1080/00018730050198152}

\bibitem{hof_lifetime}
Hof, B., de~Lozar, A., Kuik, D.J., Westerweel, J.: Repeller or attractor?
  {S}electing the dynamical model for the onset of turbulence in pipe flow.
\newblock Phys. Rev. Lett. \textbf{101}(21), 214,501 (2008).
\newblock \doi{10.1103/PhysRevLett.101.214501}

\bibitem{hof2006flt}
Hof, B., Westerweel, J., Schneider, T., Eckhardt, B.: {Finite lifetime of
  turbulence in shear flows}.
\newblock Nature \textbf{443}, 59--62 (2006)

\bibitem{itoh2006}
Itoh, K., Itoh, S.I., Diamond, P.H., Hahm, T.S., Fujisawa, A., Tynan, G.R.,
  Yagi, M., Nagashima, Y.: Physics of zonal flows.
\newblock Physics of Plasmas \textbf{13}(5), 055502 (2006).
\newblock \doi{http://dx.doi.org/10.1063/1.2178779}

\bibitem{janssen1981}
Janssen, H.: On the nonequilibrium phase transition in reaction-diffusion
  systems with an absorbing stationary state.
\newblock Zeitschrift f{\"u}r Physik B Condensed Matter \textbf{42}(2),
  151--154 (1981).
\newblock \doi{10.1007/BF01319549}

\bibitem{kadanoff1966}
Kadanoff, L.P.: Scaling laws for {I}sing models near tc.
\newblock Physics \textbf{2}, 263--272 (1966)

\bibitem{kadanoff2009more}
Kadanoff, L.P.: More is the same; phase transitions and mean field theories.
\newblock Journal of Statistical Physics \textbf{137}(5-6), 777--797 (2009)

\bibitem{kadanoff2013relating}
Kadanoff, L.P.: Relating theories via renormalization.
\newblock Studies in History and Philosophy of Science Part {B}: Studies in
  History and Philosophy of Modern Physics \textbf{44}(1), 22--39 (2013)

\bibitem{kampen1961power}
Kampen, N.v.: A power series expansion of the master equation.
\newblock Canadian Journal of Physics \textbf{39}(4), 551--567 (1961)

\bibitem{kellay2012testing}
Kellay, H., Tran, T., Goldburg, W., Goldenfeld, N., Gioia, G., Chakraborty, P.:
  Testing a missing spectral link in turbulence.
\newblock Physical Review Letters \textbf{109}(25), 254,502 (2012)

\bibitem{kim2003}
Kim, E.j., Diamond, P.H.: Zonal flows and transient dynamics of the {{L}-{H}}
  transition.
\newblock Phys. Rev. Lett. \textbf{90}, 185,006 (4 pages) (2003).
\newblock \doi{10.1103/PhysRevLett.90.185006}

\bibitem{kraichnan1959structure}
Kraichnan, R.H.: The structure of isotropic turbulence at very high reynolds
  numbers.
\newblock Journal of Fluid Mechanics \textbf{5}(04), 497--543 (1959)

\bibitem{landau1944problem}
Landau, L.D.: On the problem of turbulence.
\newblock In: Dokl. Akad. Nauk SSSR, vol.~44, pp. 339--349 (1944)

\bibitem{lemoult2016directed}
Lemoult, G., Shi, L., Avila, K., Jalikop, S.V., Avila, M., Hof, B.: Directed
  percolation phase transition to sustained turbulence in {C}ouette flow.
\newblock Nature Physics \textbf{12}, 254--258 (2016)

\bibitem{lotka1910contribution}
Lotka, A.J.: Contribution to the theory of periodic reactions.
\newblock The Journal of Physical Chemistry \textbf{14}(3), 271--274 (1910)

\bibitem{manneville2015transition}
Manneville, P.: On the transition to turbulence of wall-bounded flows in
  general, and plane {C}ouette flow in particular.
\newblock European Journal of Mechanics-B/Fluids \textbf{49}, 345--362 (2015)

\bibitem{mccomb1995theory}
McComb, W.: Theory of turbulence.
\newblock Reports on Progress in Physics \textbf{58}, 1117--1205 (1995)

\bibitem{mckane2005predator}
McKane, A.J., Newman, T.J.: Predator-prey cycles from resonant amplification of
  demographic stochasticity.
\newblock Physical Review Letters \textbf{94}(21), 218,102 (2005)

\bibitem{MEHR08}
Mehrafarin, M., Pourtolami, N.: Intermittency and rough-pipe turbulence.
\newblock Phys.\ Rev.\ E \textbf{77}, 055,304 (2008)

\bibitem{mobilia2007}
Mobilia, M., Georgiev, I.T., T\"{a}uber, U.C.: Phase transitions and
  spatio-temporal fluctuations in stochastic lattice {L}otka-{V}olterra models.
\newblock Journal of Statistical Physics \textbf{128}(1-2), 447--483 (2007).
\newblock \doi{10.1007/s10955-006-9146-3}

\bibitem{NIKU33}
Nikuradze, J.: {Stromungsgesetze in rauhen Rohren}.
\newblock VDI Forschungsheft \textbf{361}(1) (1933).
\newblock [English translation available as National Advisory Committee for
  Aeronautics, Tech. Memo. 1292 (1950). Online at:
  http://hdl.handle.net/2060/19930093938]

\bibitem{parker2014generation}
Parker, J.B., Krommes, J.A.: Generation of zonal flows through symmetry
  breaking of statistical homogeneity.
\newblock New Journal of Physics \textbf{16}(3), 035,006 (2014)

\bibitem{polyakov2015kenneth}
Polyakov, A.: {K}enneth {W}ilson in {M}oscow.
\newblock arXiv preprint arXiv:1502.03502  (2015)

\bibitem{pomeau}
Pomeau, Y.: Front motion, metastability and subcritical bifurcations in
  hydrodynamics.
\newblock Physica \textbf{23D}, 3--11 (1986)

\bibitem{pomeau2016long}
Pomeau, Y.: The long and winding road.
\newblock Nature Physics \textbf{12}, 198--199 (2016)

\bibitem{prost2009generalized}
Prost, J., Joanny, J.F., Parrondo, J.: Generalized fluctuation-dissipation
  theorem for steady-state systems.
\newblock Physical Review Letters \textbf{103}(9), 90,601 (2009)

\bibitem{renshaw1993modelling}
Renshaw, E.: Modelling biological populations in space and time, vol.~11.
\newblock Cambridge University Press (1993)

\bibitem{REYN83}
Reynolds, O.: {An Experimental Investigation of the Circumstances which
  determine whether the Motion of Water shall be Direct or Sinuous and the Law
  of Resistance in Parallel Channel}.
\newblock Philos. Trans. R. Soc. London \textbf{174}, 935 (1883)

\bibitem{reynolds}
Reynolds, O.: An experimental investigation of the circumstances which
  determine whether the motion of water shall be direct or sinuous, and of the
  law of resistance in parallel channels.
\newblock Phil. Trans. Roy. Soc. A \textbf{174}, 935--982 (1883)

\bibitem{ruelle2014non}
Ruelle, D.: Non-equilibrium statistical mechanics of turbulence.
\newblock Journal of Statistical Physics \textbf{157}, 205--218 (2014)

\bibitem{ruelle1971nature}
Ruelle, D., Takens, F.: On the nature of turbulence.
\newblock Communications in mathematical physics \textbf{20}(3), 167--192
  (1971)

\bibitem{sano2016universal}
Sano, M., Tamai, K.: A universal transition to turbulence in channel flow.
\newblock Nature Physics \textbf{12}, 249--253 (2016)

\bibitem{schmitz2012}
Schmitz, L., Zeng, L., Rhodes, T.L., Hillesheim, J.C., Doyle, E.J., Groebner,
  R.J., Peebles, W.A., Burrell, K.H., Wang, G.: Role of zonal flow
  predator-prey oscillations in triggering the transition to {H}-mode
  confinement.
\newblock Phys. Rev. Lett. \textbf{108}, 155,002 (5 pages) (2012).
\newblock \doi{10.1103/PhysRevLett.108.155002}

\bibitem{schneider2008lifetime}
Schneider, T., Eckhardt, B.: {Lifetime statistics in transitional pipe flow}.
\newblock Physical Review E \textbf{78}(4), 46,310 (2008)

\bibitem{seifert2012stochastic}
Seifert, U.: Stochastic thermodynamics, fluctuation theorems and molecular
  machines.
\newblock Reports on Progress in Physics \textbf{75}(12), 6001 (2012)

\bibitem{dpturb2016}
Shih, H.Y., Goldenfeld, N.: Extreme value statistics and critical exponents at
  the laminar-turbulence transition in pipes (2016).
\newblock Unpublished

\bibitem{shih2016ecological}
Shih, H.Y., Hsieh, T.L., Goldenfeld, N.: Ecological collapse and the emergence
  of travelling waves at the onset of shear turbulence.
\newblock Nature Physics \textbf{12}, 245--248 (2016)

\bibitem{sipos2011directed}
Sipos, M., Goldenfeld, N.: Directed percolation describes lifetime and growth
  of turbulent puffs and slugs.
\newblock Physical Review E \textbf{84}(3), 035,304 (2011)

\bibitem{sivashinsky1985negative}
Sivashinsky, G., Yakhot, V.: Negative viscosity effect in large-scale flows.
\newblock Physics of Fluids (1958-1988) \textbf{28}(4), 1040--1042 (1985)

\bibitem{song2014deterministic}
Song, B., Hof, B.: Deterministic and stochastic aspects of the transition to
  turbulence.
\newblock Journal of Statistical Mechanics: Theory and Experiment
  \textbf{2014}(2), P02,001 (2014)

\bibitem{SREE05}
Sreenivasan, K.R., Eyink, G.L.: {S}am {E}dwards and the turbulence theory.
\newblock In: P.~Goldbart, N.~Goldenfeld, D.~Sherrington (eds.) Stealing the
  Gold: a celebration of the pioneering physics of {S}am {E}dwards, pp. 66--85.
  Oxford University Press (2005)

\bibitem{STRI23}
Strickler, A.: Beitrage zur frage der geschwindigkeitsformel und der
  rauhigkeitszahlen fur strome, kanale und geschlossene leitungen (1923).
\newblock Mitteilungen des Eidgen{\" o}ssischen Amtes f{\" u}r Wasserwirtschaft
  16, Bern, Switzerland. ~Translated as \lq\lq Contributions to the question of
  a velocity formula and roughness data for streams, channels and closed
  pipelines." by T. Roesgan and W. R. Brownie, Translation T-10, W. M. Keck Lab
  of Hydraulics and Water Resources, Calif. Inst. Tech., Pasadena, Calif.
  January 1981

\bibitem{tauber2012}
T{\"a}uber, U.C.: Population oscillations in spatial stochastic
  {L}otka--{V}olterra models: a field-theoretic perturbational analysis.
\newblock Journal of Physics A: Mathematical and Theoretical \textbf{45}(40),
  405,002 (2012)

\bibitem{tel2008cts}
T{\'e}l, T., Lai, Y.: {Chaotic transients in spatially extended systems}.
\newblock Physics Reports \textbf{460}(6), 245--275 (2008)

\bibitem{TRAN10}
Tran, T., Chakraborty, P., Guttenberg, N., Prescott, A., Kellay, H., Goldburg,
  W., Goldenfeld, N., Gioia, G.: Macroscopic effects of the spectral structure
  in turbulent flows.
\newblock Nature Phys. \textbf{4}, 438--441 (2010)

\bibitem{van2011stochastic}
Van~Kampen, N.: Stochastic Processes in Physics and Chemistry.
\newblock Elsevier (2011)

\bibitem{volterra1927variazioni}
Volterra, V.: Variazioni e fluttuazioni del numero d'individui in specie
  animali conviventi.
\newblock C. Ferrari (1927)

\bibitem{WIDO65}
Widom, B.: Equation of state in the neighbourhood of the critical point.
\newblock J. Chem. Phys. \textbf{43}, 3898--3905 (1965)

\bibitem{widom2011laboring}
Widom, B.: Laboring in the vineyard of physical chemistry.
\newblock Annual Review of Physical Chemistry \textbf{62}, 1--18 (2011)

\bibitem{WK09}
Willis, A.P., Kerswell, R.R.: Turbulent dynamics of pipe flow captured in a
  reduced model: puff relaminarisation and localised `edge' states.
\newblock J.\ Fluid Mech. \textbf{619}, 213--233 (2009)

\bibitem{wilson1971renormalization}
Wilson, K.G.: Renormalization group and critical phenomena. {I}.
  renormalization group and the {K}adanoff scaling picture.
\newblock Physical Review B \textbf{4}, 3174--3183 (1971)

\bibitem{wilson1983renormalization}
Wilson, K.G.: The renormalization group and critical phenomena.
\newblock Reviews of Modern Physics \textbf{55}, 583--600 (1983)

\bibitem{wyld1961formulation}
Wyld, H.W.: Formulation of the theory of turbulence in an incompressible fluid.
\newblock Annals of Physics \textbf{14}, 143--165 (1961)

\bibitem{xu2011}
Xu, G.S., Wan, B.N., Wang, H.Q., Guo, H.Y., Zhao, H.L., Liu, A.D., Naulin, V.,
  Diamond, P.H., Tynan, G.R., Xu, M., Chen, R., Jiang, M., Liu, P., Yan, N.,
  Zhang, W., Wang, L., Liu, S.C., Ding, S.Y.: First evidence of the role of
  zonal flows for the {L}-{H} transition at marginal input power in the east
  tokamak.
\newblock Phys. Rev. Lett. \textbf{107}, 125,001 (5 pages) (2011).
\newblock \doi{10.1103/PhysRevLett.107.125001}

\bibitem{yakhot1986renormalization}
Yakhot, V., Orszag, S.A.: Renormalization group analysis of turbulence. {I}.
  basic theory.
\newblock Journal of scientific computing \textbf{1}(1), 3--51 (1986)

\bibitem{zuniga2014spectral}
Z{\'u}{\~n}iga~Zamalloa, C., Ng, H.C.H., Chakraborty, P., Gioia, G.: Spectral
  analogues of the law of the wall, the defect law and the log law.
\newblock Journal of Fluid Mechanics \textbf{757}, 498--513 (2014)

\end{thebibliography}
\end{document}